
%
%
\input phyzzx
%
%
%
%
%
%
%
%
%
%
%
%
\catcode`@=11

\font\seventeencp=cmcsc10 scaled\magstep3

\def\OFFSET{\voffset=1.2truein\hoffset=.8truein}
\Pubnum={\rm CERN$-$TH.\the\pubnum }
\def\title#1{\vskip\frontpageskip\vskip .50truein
     \titlestyle{\seventeencp #1} \vskip\headskip\vskip\frontpageskip
     \vskip .2truein}
\def\author#1{\vskip .27truein\titlestyle{#1}\nobreak}

\def\p@bblock{\begingroup \tabskip=\hsize minus \hsize
   \baselineskip=1.5\ht\strutbox \topspace-2\baselineskip
   \halign to\hsize{\strut ##\hfil\tabskip=0pt\crcr
   \the \Pubnum\cr}\endgroup}
\def\makefootline{\iffrontpage\vskip .27truein\line{\the\footline}
                 \vskip -.1truein\line{\the\date\hfil}
              \else\line{\the\footline}\fi}
\paperfootline={\iffrontpage \the\Pubnum\hfil\else\hfil\fi}
\paperheadline={\iffrontpage\hfil
                \else\twelverm\hss $-$\ \folio\ $-$\hss\fi}
\newif\ifmref  
\newif\iffref  
\def\xrefsend{\xrefmark{\count255=\referencecount
\advance\count255 by-\lastrefsbegincount
\ifcase\count255 \number\referencecount
\or \number\lastrefsbegincount,\number\referencecount
\else \number\lastrefsbegincount-\number\referencecount \fi}}
\def\xrefsdub{\xrefmark{\count255=\referencecount
\advance\count255 by-\lastrefsbegincount
\ifcase\count255 \number\referencecount
\or \number\lastrefsbegincount,\number\referencecount
\else \number\lastrefsbegincount,\number\referencecount \fi}}
\def\xREFNUM#1{\space@ver{}\refch@ck\firstreflinetrue%
\global\advance\referencecount by 1
\xdef#1{\xrefend}}
\def\xrefend{\xrefmark{\number\referencecount}}
\def\xrefmark#1{[{#1}]}
\def\xRef#1{\xREFNUM#1\immediate\write\referencewrite%
{\noexpand\refitem{#1}}\begingroup\obeyendofline\rw@start}%
\def\xREFS#1{\xREFNUM#1\global\lastrefsbegincount=\referencecount%
\immediate\write\referencewrite{\noexpand\refitem{#1}}%
\begingroup\obeyendofline\rw@start}
\def\rrr#1#2{\relax\ifmref{\iffref\xREFS#1{#2}%
\else\xRef#1{#2}\fi}\else\xRef#1{#2}\xrefend\fi}
\def\multref#1#2{\mreftrue\freftrue{#1}%
\freffalse{#2}\mreffalse\xrefsend}
\referencecount=0
%
\space@ver{}\refch@ck\firstreflinetrue%
\immediate\write\referencewrite{}%
\begingroup\obeyendofline\rw@start{}%
\def\plb#1({Phys.\ Lett.\ $\underline  {#1B}$\ (}
\def\nup#1({Nucl.\ Phys.\ $\underline {B#1}$\ (}
\def\plt#1({Phys.\ Lett.\ $\underline  {B#1}$\ (}
\def\cmp#1({Comm.\ Math.\ Phys.\ $\underline  {#1}$\ (}
\def\prp#1({Phys.\ Rep.\ $\underline  {#1}$\ (}
\def\prl#1({Phys.\ Rev.\ Lett.\ $\underline  {#1}$\ (}
\def\prv#1({Phys.\ Rev. $\underline  {D#1}$\ (}
\def\und#1({            $\underline  {#1}$\ (}
\message{ by W.L.}
\everyjob{\input offset }
\catcode`@=12

\def\ie{{\sl i.e.}}
\let\it=\sl

\def\KIYA{\rrr\KIYA{K. Kikkawa and M. Yamasaki, \plb149 (1984) 357.}}

\def\SASE{\rrr\SASE{N. Sakai and I. Senda, Prog. Theor. Phys.
75 (1986) 692.}}

\def\ZUC{\rrr\ZUC{R. Zucchini, \nup350 (1991) 111.}}

\def\FIQS{\rrr\FIQS{A. Font, L.E. Iba\~nez, F. Quevedo and A. Sierra,
\nup337 (1990) 119.}}

\def\CLS{\rrr\CLS{P. Candelas, M. Lynker and R. Schimmrigk,
\nup341 (1990) 402.}}

\def\TAYLOR{\rrr\TAYLOR{T.R. Taylor, \plt252 (1990) 59.}}

\def\LMNA{\rrr\LMNA{J. Lauer, J. Mas and H.P. Nilles,
\nup351 (1991) 353.}}

\def\BCF{\rrr\BCF{R. Barbieri, E. Cremmer and S.Ferrara,\plb163 (1985)
143.}}

\def\POLONY{\rrr\POLONY{J. Polony, Budapest preprint K-FKI-1977-83
(1977).}}

\def\CHSW{\rrr\CHSW{P. Candelas, G. Horowitz, A. Strominger and
E. Witten, \nup258 (1985) 46.}}

\def\FEST{\rrr\FEST{S. Ferrara and A. Strominger,
in Proc. Texas A and M String 89 Workshop.}}

\def\FIQ{\rrr\FIQ{A. Font, L. Iba\~nez and F. Quevedo, \nup345 (1990)
389.}}

\def\GHMR{\rrr\GHMR{D. Gross, J. Harvey, E. Martinec and R. Rohm,
\nup256 (1985) 253.}}

\def\DIX{\rrr\DIX{L. Dixon in {\it Proceedings of the 1987 ICTP
Summer Workshop in High Energy Physics and Cosmology,} Trieste.}}

\def\LLS{\rrr\LLS{W. Lerche, D. L\"ust and A.N. Schellekens,
\nup287 (1987) 477.}}

\def\KLT{\rrr\KLT{H. Kawai, D.C. Lewellen and S.-H.H. Tye,
\nup288 (1987) 1.}}

\def\ABK{\rrr\ABK{I. Antoniadis, C.P. Bachas and C. Kounnas,
\nup289 (1987) 87.}}

\def\HAVA{\rrr\HAVA{S. Hamidi and C. Vafa, \nup279 (1987) 465.}}

\def\LVW{\rrr\LVW{W. Lerche, C. Vafa and N. Warner, \nup324 (1989) 427.}}

\def\GIPO{\rrr\GIPO{A. Giveon and M. Porrati, \plt246 (1990) 54.}}

\def\ANTO{\rrr\ANTO{I. Antoniadis, \plt246 (1990) 377.}}

\def\GISM{\rrr\GISM{A. Giveon and D.-J. Smit,
{\it ``Symmetries on the moduli space of (2,2) superstring vacua'',}
LBL-28864, UCB-PTH--90/15 (1990).}}

\def\CVEMOLOV{\rrr\CVEMOLOV{M. Cveti\v c, J. Molera and B. Ovrut,
\plt248 (1990) 83.}}

\def\ASLUE{\rrr\ASLUE{P.S. Aspinwall and C.A. L\"utken,
{\it ``Quantum algebraic geometry of superstring compactification'',}
preprint (1990).}}

\def\PASA{\rrr\PASA{
J. C. Pati and A. Salam, \prv10 (1974) 275.}}

\def\MOSEN{\rrr\MOSEN{
G. Senjanovi\' c and R. N. Mohapatra, \prv12
(1975) 1502.}}

\def\CVETIC{\rrr\CVETIC{M. Cveti\v c, in the
{\it Proceedings of  the Conference on
 Elementary Particle Physics}, Tuscaloosa, AL, November 1989
(World Scientific 1990, L.~Clavelli et al, eds.).}}

\def\LT{\rrr\LT{D. L\"ust and T.R. Taylor, \plt253 (1991) 335.}}

\def\GRKL{\rrr\GRKL{D.J. Gross and I.R. Klebanov, \nup344 (1990) 475.}}

\def\ALBA{\rrr\ALBA{E. Alvarez and J.L.F. Barbon, {\it ``Spacetime
R-duality in discretized string theories'',} preprint FTUAM/31-90
(1990).}}

\def\SENDA{\rrr\SENDA{I. Senda, \nup318 (1989) 211.}}

\def\SAT{\rrr\SAT{B. Sathiapalan, \prl58 (1987) 1597.}}

\def\NSSW{\rrr\NSSW{V.P. Nair, A. Shapere, A. Strominger and F. Wilczek,
      \nup287 (1987) 402.}}

\def\DVV{\rrr\DVV
{R.~Dijkgraaf, E.~Verlinde and H.~Verlinde, \cmp115 (1988) 649;
{\it ``On moduli
  spaces of conformal field theories with $c\geq 1$'',}
  preprint THU-87/30.}}

\def\KAPLU{\rrr\KAPLU{V. Kaplunovsky, \nup307 (1988) 145.}}

\def\MINAH{\rrr\MINAH{J. Minahan, \nup298 (1988) 36.}}

\def\TV{\rrr\TV{T.R. Taylor and G. Veneziano, \plt212 (1988) 147.}}

\def\FLST{\rrr\FLST{S. Ferrara,
 D. L\"ust, A. Shapere and S. Theisen,
   \plt225 (1989) 363.}}

\def\MOOV{\rrr\MOOV{J. Molera and B. Ovrut, \prv40 (1989) 1146.}}

\def\GRPL{\rrr\GRPL{B.R. Greene and M.R. Plesser, \nup338 (1990) 15.}}

\def\AO{\rrr\AO{E. Alvarez and M. Osorio, \prv40 (1989) 1150.}}

\def\LMN{\rrr\LMN{J. Lauer, J. Mas and H.P. Nilles, \plt226 (1989) 251.}}

\def\LLW{\rrr\LLW{
W. Lerche, D. L\"ust and N.P. Warner, \plt231 (1989) 417.}}

\def\CMLN{\rrr\CMLN
{E.J. Chun, J. Mas, J. Lauer and H.P. Nilles, \plt233 (1989) 141.}}

\def\FLT{\rrr\FLT
{S. Ferrara, D. L\"ust and S. Theisen, \plt233 (1989) 147.}}

\def\ILLT{\rrr\ILLT{L. Iba\~nez, W. Lerche, D. L\"ust and S. Theisen,
   {\it ``Some considerations about the stringy Higgs effect''},
   CERN-TH.5654/90, CALT-68-1620, to appear in Nucl. Phys. B.}}

\def\FLTA{\rrr\FLTA{S. Ferrara, D. L\"ust and S. Theisen,
    \plt242 (1990) 39.}}

\def\SCHW{\rrr\SCHW{
J.H. Schwarz, {\it ``Space-time duality in string theory''},
preprint CALT--68--1581.}}

\def\DHVW{\rrr\DHVW{L. Dixon, J. Harvey, C.~Vafa and E.~Witten,
\nup261 (1985) 651; \nup274 (1986) 285.}}

\def\CLOB{\rrr\CLOB
{M. Cveti\v c, B. Ovrut and J. Louis, \prv40 (1989) 684.}}

\def\WI{\rrr\WI{E. Witten, \plb155 (1985) 151.}}

\def\FKP{\rrr\FKP
{S. Ferrara, C. Kounnas and M. Porrati, \plt181 (1986) 263.}}

\def\DKL{\rrr\DKL
{L. Dixon, V. Kaplunovsky and J. Louis, \nup329 (1990) 27.}}

\def\CLOA{\rrr\CLOA{M. Cveti\v c, J. Louis and B. Ovrut,
\plt206 (1988) 227.}}

\def\INQ{\rrr\INQ{L.E. Iba\~nez, H.P.~Nilles and F.~Quevedo,
   \plt187 (1987) 25; \plt192 (1987) 332.}}

\def\FINQ{\rrr\FINQ{A. Font, L.E. Iba\~nez, H.P. Nilles and F. Quevedo,
\nup307 (1988) 109.}}

\def\FEPO{\rrr\FEPO{S. Ferrara and M. Porrati, \plt 216 (1989) 289.}}

\def\CMO{\rrr\CMO{M. Cveti\v c, J. Molera and B. Ovrut,
\prv40 (1989) 1140.}}

\def\DINA{\rrr\DINA{J.P. Derendinger, L.E. Iba\~nez and H.P. Nilles,
\nup267 (1986) 365.}}

\def\DHS{\rrr
\DHS{M. Dine, P. Huet and N. Seiberg, \nup322 (1989) 301.}}

\def\GRV {\rrr
\GRV{A. Giveon, E. Rabinovici and G. Veneziano,
    \nup322 (1989) 167.}}

\def\SHWI{\rrr
\SHWI{A. Shapere and F. Wilzcek, \nup320 (1989) 669.}}

\def\DIN{\rrr\DIN{J.P. Derendinger, L.E. Ib\'a\~nez and H.P. Nilles,
     \plb155 (1985) 65.}}

\def\DRSW{\rrr\DRSW{M. Dine, R. Rohm, N. Seiberg and E. Witten,
   \plb156 (1985) 55.}}

\def\bigpaper{\rrr\bigpaper{M. Cveti\v c, A. Font, L.E. Ib\'a\~nez,
D. L\"ust and
F. Quevedo, \sl ``Target Space Duality, Supersymmetry Breaking and the
Stability of Classical String Vacua,'' Nucl. Phys. \bf B361 \rm (1991)
194.}}

\def\CQRI{\rrr\CQRI{M. Cveti\v c, F. Quevedo, and S.-J. Rey, \sl ``Target
Space Duality and Stringy Domain Walls,''
\rm UPR--0445--T to appear in
Phys. Rev. Lett.}}

\def\CQRII{\rrr\CQRII{M. Cveti\v c, F. Quevedo, and S.-J. Rey, \sl
``Stringy Topological Defects,'' in preparation; M. Cveti\v c
`` Topological Defects in the Moduli Sector of String Theory '',
UPR-485, Talk presented at Srings'91 Workshop, Stony Brook.}}

\def\vafastring{\rrr\vafastring{B. Greene, A. Shapere, C. Vafa and S.T. Yau,
Nucl. Phys. \bf B337 \rm (1990) 1.}}

\def\LEHNER{\rrr\LEHNER{J. Lehner, \sl``Discontinuous Groups and
Automorphic Functions", \rm ed. by the American Mathematical Society,
(1964).}}

\def\DFKZ{\rrr\DFKZ{J.P. Derendinger, S. Ferrara, C. Kounnas, and F.
Zwirner, CERN--TH--6004; J.P. Derendinger,
these proceedings.}}

\def\CO{\rrr\CO{G. Cardoso and B. Ovrut, UPR--0464--T and
 UPR--0481--T.}}

\def\LOU{\rrr\LOU{J. Louis, these proceedings.}}

\def\F{\rrr\F{S. Ferrara, preprint UCLA/90/TEP/20 (1990).}}

\def\FGN{\rrr\FGN{S. Ferrara, L. Girardello and H.P. Nilles,
   \plb125 (1983) 457. }}

\def\CFGVP{\rrr\CFGVP{E. Cremmer, S. Ferrara, L. Girardello and
A. Van Proeyen, \nup212 (1983) 413. }}

\def\IN{\rrr\IN{L.E. Ib\`a\~nez and H.P. Nilles, \plb169 (1986) 354.}}

\def\CLMR{\rrr\CLMR{J.A. Casas, Z. Lalak, C. Mu\~noz and G.G. Ross,
      \nup347 (1990) 243.}}

\def\DIXON{\rrr\DIXON{
L. Dixon, {\it ``Supersymmetry breaking in string theory'',}
SLAC preprint 5229 (1990).}}

\def\KRAS{\rrr\KRAS{N.V. Krasnikov, \plt193 (1987) 37.}}

\def\HPN{\rrr\HPN{H.P. Nilles, \plb115 (1982) 193; \nup217 (1983) 366.}}

\def\DKLA{\rrr\DKLA{L. Dixon, V. Kaplunovsky and J. Louis,
{\sl ``Moduli-dependence of string loop corrections to gauge
coupling constants'',}
Nucl. Phys. \rm 355B \ (1991) 649.}}

\def\FERRARA{\rrr\FERRARA{S. Ferrara, {\sl ``Superstring effective field
theories on (2,2) vacua'',} CERN-TH.5873/90 (1990).}}

\def\STR{\rrr\STR{A. Strominger, Comm. Math. Phys. 133 (1990)
163.}}

\def\CAF{\rrr\CAF{L. Castellani, R. D'Auria and S. Ferrara,
Class. Quantum Grav. 7 (1990) 1767.}}

\def\CAN{\rrr\CAN{P. Candelas, X.C. De La Ossa, P.S. Green and L. Parkes,
{\it ``A pair of Calabi-Yau Manifolds as an exactly soluble
superconformal theory'',} UTTG-25-1990 (1990).}}

\def\FILQ{\rrr\FILQ{A. Font, L.E. Ib\'a\~nez, D. L\"ust and
F. Quevedo, \plt245 (1990) 401.}}

\def\FILQA{\rrr\FILQA{A. Font, L.E. Ib\'a\~nez, D. L\"ust and
F. Quevedo, \plt249 (1990) 35.}}

\def\FMTV{\rrr\FMTV{S. Ferrara, N. Magnoli, T.R. Taylor and
G. Veneziano, \plt245 (1990) 409.}}

\def\GT{\rrr\GT{M.K. Gaillard, these proceedings; T. Taylor, these
           proceedings.}}

\def\NIOL{\rrr\NIOL{H.P. Nilles and M. Olechowski, \plt248 (1990) 268.}}

\def\BIN{\rrr\BIN{P. Binetruy and M.K. Gaillard, \plt253 (1991) 119.}}

\def\DUFF{\rrr\DUFF{M. Duff, \nup335 (1990) 610.}}

\def\CREMMER{\rrr\CREMMER{E. Cremmer, S. Ferrara, L. Girardello
and A. Van Proeyen, \nup212 (1983) 413. }}

\def\DSWW{\rrr\DSWW{M. Dine, N. Seiberg, X. Wen and E. Witten,
\nup278 (1986) 769, \nup289 (1987) 357. }}

\def\DFMS{\rrr\DFMS{L. Dixon, D. Friedan, E. Martinec and
S. Shenker, \nup282 (1987) 13.}}

\def\PSS{\rrr\PSS{Y. Park, M. Srednicki and
A. Strominger, \plt244 (1990) 393.}}

\def\REY{\rrr\REY{S.-J. Rey, \prv43 (1991) 526.}}

\def\FORD{\rrr\FORD{R. Ford, {\it Automorphic functions}, Chelsea
Publishing Company, New York.}}

\def\BRFR{\rrr\BRFR{P. Breitenlohner and D. Freedman, \plb115
(1982) 197.}}

\def\CHOI{\rrr\CHOI{K. Choi and J.E. Kim, Phys. Rev. Lett. \bf 55 \rm
(1985) 2637; J.A. Casas and G.G. Ross, Phys. Lett. \bf B198 \rm (1987) 461.}}

\def\SCP{\rrr\SCP{E. Witten, \plb149 (1984) 351; \nextline
J.E. Kim, \plb154 (1985) 393; \plb165 (1985) 71;\plb207 (1988) 434; \nextline
E. Witten and X. Wen, \plb166 (1986) 397;\nextline
G. Lazarides, C. Panagiotakopoulos and Q. Shafi Phys.Rev.Lett.56 (1986)
432; \nextline
J.A. Casas and G.G. Ross, \plb192  (1987) 119; \nextline
J. Lopez and D.V. Nanopoulos, \plb245  (1990) 111.}}

\def\SW{\rrr\SW{E. Witten and A. Strominger, Commun. Math, Phys. 101 (1985)
341}}

\def\KL{\rrr\KL{V. Kaplunovsky and J. Louis, to appear.}}

\def\NSV{\rrr\NSV{K. Narain, M. Sarmadi and C. Vafa, \nup288 (1987) 551. }}

\def\SVA{\rrr\SVA{M. Shifman and A. Vainshtein, \nup277 (1986) 456.}}

\def\SVB{\rrr\SVB{M. Shifman and A. Vainshtein, preprint TPI-Minn-91/4-T
(1991).}}

\def\KUGO{\rrr\KUGO{T. Kugo and S. Uehara, \nup222 (1983) 125.}}

\def\O{\rrr\O{G. Lopez Cardoso and B. Ovrut, prep. UPR-0464T (1991).}}

\def\ORO{\rrr\ORO{A. Font, L.E. Ib\'a\~nez, H.P. Nilles and F. Quevedo,
\plb210 (1988) 101;\nextline
A. Font, L.E. Ib\'a\~nez, F. Quevedo and A. Sierra, \nup331 (1990) 421.}}

\def\ROFLIP{\rrr\ROFLIP{B. Greene, K. Kirklin, P. Miron and G.G. Ross,
\nup278 (1986) 668; \nup292 (1987) 606; \nextline
I. Antoniadis, J. Ellis, J. Hagelin and D.V. Nanopoulos, \plb205 (1988) 459;
\plb231 (1989) 65. }}

\def\SVZ{\rrr\SVZ{M. Shifman, V. Vainshtein and V. Zakharov, \nup166 (1980)
493. }}

\def\CFILQ{\rrr\CFILQ{M. Cveti\v c , A. Font, L.E. Ib\'a\~nez, D. L\"ust
and F. Quevedo, Nucl.Phys.\bf B361 \rm (1991) 194. }}

\def\KIM {\rrr\KIM{J.E. Kim, Phys.Rev.Lett. 43 (1979) 103.}}

\def\axion{\rrr\axion{J.E. Kim, Phys. Rev. Lett. \bf 43 \rm (1979) 103:
M. Dine,
W. Fischler and M. Srednicki, Phys. Lett. \bf 104B \rm (1981) 199
and
Nucl. Phys. \bf B189 \rm (1981) 575;
 M.B. Wise, H. Georgi and S.L. Glashow,
Phys. Rev. Lett. \bf 47 \rm (1981) 402: a good review is by J.E. Kim,
Phys. Rep. \bf 150 \rm (1987) 1.}}

\def\original{\rrr\original{See for instance, S. Coleman,
\sl ``Aspects of Symmetry,'' \rm
Chap.7, Cambridge Univ. Press. (1985) or R. Rajaraman \sl``Solitons and
Instantons''\rm\ North-Holland (1987).}}

\def\ovrut{\rrr\ovrut{B. Ovrut and S. Thomas, \sl ``Instantons in Antisymmetric
Tensor Theories in Four Dimensions,'' \rm  UPR-465-T preprint (January,
1991).}}

\def\vilen{\rrr\vilen{P. Sikivie, Phys. Rev. Lett. \bf 48 \rm (1982) 1156;
G. Lazarides and Q. Shafi, Phys. Lett. \bf 115B\rm (1982) 21;
for a review, see
A. Vilenkin, Phys. Rep. \bf 121 \rm (1985) 263.}}

\def\worldsheetinstanton{\rrr\worldsheetinstanton{M. Dine, N. Seiberg,
X. Wen and E. Witten,
Nucl. Phys. \bf B278 \rm (1986) 769  and   Nucl. Phys. \bf B289 \rm
(1987) 319.}}

\def\modularform{\rrr\modularform{B. Schoeneberg,
\sl ``Elliptic Modular Functions,'' \rm
Springer, Berlin-Heidelberg (1970);
J. Lehner, \sl ``Discontinuous Groups and
Automorphic Functions,'' \rm
 ed. by the American Mathematical Society,
(1964).}}

\def\giveon{\rrr\giveon{A. Giveon, N. Malkin and E. Rabinovici, Phys. Lett.
 \bf B238 \rm (1990) 57.}}

\def\vafa{\rrr\vafa{
P. Fendley, S. Mathur, C. Vafa and N.P. Warner, Phys. Lett.
\bf B243\rm (1990) 257.}}

\def\rey{\rrr\rey{
S.-J. Rey, \sl ``Axionic String Instantons and Their
Low-Energy Implications,'' \rm Invited Talk at Tuscaloosa Workshop on
Particle Theory and Superstrings, ed. L. Clavelli and B. Harm, World
Scientific Pub., (November, 1989): Phys. Rev. \bf D43\rm (1991) 526.}}

\def\WO{\rrr\WO{E. Witten and D. Olive, Phys. Lett. \bf 78B\rm (1978) 97.}}

\def\domainansatz{\rrr\domainansatz{A. Vilenkin, Phys. Lett. \bf 133B
\rm (1983) 177:
J. Ipser and P. Sikivie, Phys. Rev. \bf D30 \rm (1984) 712.}}

\def\positiveenergy{\rrr\positiveenergy{E. Witten, Comm. Math. Phys.
\bf 80 \rm (1981) 381.}}

\def\ours{\rrr\ours{M. Cveti\v c, S. Griffies, and S.-J. Rey,
`` Physical Implications of Stringy Domain Walls,'' UPR-741 preprint.}}

\def\falsevacuum{\rrr\falsevacuum{S. Coleman, Phys. Rev.
\bf D15 \rm (1977) 2929;
C. Callan and S. Coleman, Phys. Rev. \bf D16 \rm (1977) 1762.}}

\def\sqr#1#2{{\vcenter{\vbox{\hrule height.#2pt
          \hbox{\vrule width.#2pt height#1pt \kern#1pt
                \vrule width.#2pt}
                \hrule height.#2pt}}}}
\def\square{\mathchoice\sqr{8}7\sqr{8}7\sqr{8}7\sqr{8}7}
\def\OFFSET{\hoffset=6.pt\voffset=40.pt}

\def\to{\rightarrow}
\def\o{\over}

\def\p0{\phi_0}

\OFFSET

\catcode`@=12
\newtoks\Pubnumtwo
\newtoks\Pubnumthree
\catcode`@=11
\def\p@bblock{\begingroup\tabskip=\hsize minus\hsize
   \baselineskip=1.5\ht\strutbox\topspace-2\baselineskip
   \halign to \hsize{\strut ##\hfil\tabskip=0pt\crcr
   \the\Pubnum\cr  
   \the\pubtype\cr}\endgroup}
\Pubnum={UPR--0484--T}
\date{September 1991}
\pubtype={}
\titlepage\singlespace\nopagenumbers
\vskip -1.truein
\title{Implications of Target Space Duality:\break
   {\fourteenpoint Supersymmetry Breaking, Stability of String Vacua,\break
                 and Stringy Domain Walls
     \foot{Invited talk presented at  PASCOS '91, March 25--30, 1991,
                Northeastern University, Boston, MA}}}
\vskip -1.truein
\author{M. Cveti\v c}
\vskip 0.3truein
\centerline{Department of Physics}
\centerline{University of Pennsylvania}
\centerline{Philadelphia, PA 19104-6396}
\abstract\noindent\nobreak
Based on the assumption that the target space duality ($T\rightarrow
1/T$) is preserved even nonperturbatively, the properties of static
string vacua are studied. A discussion of the effective four-dimensional
supergravity action based on target-space modular symmetry $SL(2,{\bf
Z})$ is presented.  The nonperturbative superpotential removes  the
vacuum degeneracy with respect to the compactification modulus ($T$)
generically breaks supersymmetry with negative cosmological constant.
Charged matter fields get negative $(mass)^2$ signalling an additional
instability of string vacuum and the blowing up of orbifold
singularities.  In addition for a class of modularly invariant
potentials topologically stable stringy domain walls of nontrivial
compaction modulus field  configuration are found.  They are
supersymmetric solutions, thus saturating the Bogomolnyi bound.  Their
physical implications are discussed.
\vskip -1.0cm
\endpage
%
\chap{Introduction}

Non-perturbative effects are believed to play a crucial role
in low-energy string theory. They are expected to provide the
mechanism for supersymmetry breaking, to solve the dilaton
problem by generating a non-trivial potential for that field, to
lift the vacuum degeneracy existent in string theories to all
orders of perturbation theory and possibly also to give some light
in solving the cosmological constant problem.
Unfortunately, despite many efforts, a non-perturbative formulation
of string theory is not available at the moment and these fundamental
questions cannot yet be addressed from first principles.

Nevertheless,
it is possible to use the known symmetries in perturbation
theory to obtain general information about how the non-perturbative
effects could appear in the low-energy effective action. Recently,
under the assumption that target space duality
($R\rightarrow \alpha'/R$ \multref\KIYA{\SASE\SAT\NSSW\AO\MOOV\DUFF
\SCHW})
and its generalization
(the modular group $SL(2,{\bf Z})$ \multref\GRV{\SHWI\DVV\LLW})
is an exact symmetry of string theory,
strong constraints have been found \FLST\ on the general form of the
non-perturbative four-dimensional effective supergravity action,
providing a link between the superpotential and the theory of
modular functions.
It has also been realized \multref\FILQ{\FMTV\NIOL\BIN\GT}\
that very interesting consequences appear
after reconciling the well-known
gaugino condensation mechanism \multref\FGN{\HPN\DIN\DRSW}\
for supersymmetry breaking with the
requirement of target space modular invariance.
As a general result \FILQ,\FMTV,
the effective potential for the modulus field, representing the
``size'' of the internal space, is such that the theory dynamically
prefers to be
compactified, therefore lifting the vacuum degeneracy existent for this
field in string perturbation theory.

Guided by the general theory of $SL(2,{\bf Z})$ automorphic
functions, one can extend the analysis \bigpaper\
to the most general, modular invariant superpotential, independent of
the non-perturbative source  that generates it.

The second way to generalize the results, is to consider the
inclusion of ``matter fields''. In the case of orbifold
compactifications they correspond to twisted moduli (blowing-up modes),
untwisted and twisted matter. For simplicity they were set to their
vanishing vacuum expectation values
($vev's$) in previous analysis.
It is possible to
investigate the effective action perturbatively around their
zero $vev$.

The third intriguing direction is to study the existence of topological
defects \multref\CQRI{\CQRII}\ based on duality invariant nonperturbative
superpotential. They can have important cosmological
implications for the theory.  Recently cosmic string solutions (stringy
cosmic strings) \vafastring\ were found in the perturbative string theory.
It will be shown that there is a domain wall solution for a class of
duality preserving nonperturbative potentials.

Most of the work presented here is based on two papers: \bigpaper\ and
\CQRI.
This paper is organized as follows.
In the next section we discuss the form of $N=1$ supersymmetric
low-energy effective string actions which are invariant under target
space duality symmetry,
restricted $SL(2,{\bf Z})$.
We present the most general non-perturbative
effective supergravity action for the single modulus $T$,
invariant under this symmetry, and discuss its vacuum structure
which generically breaks space-time supersymmetry
with and (for particular choices of the superpotential) without
creating a cosmological constant. We briefly mention the analysis
to the case of three moduli.
In section three, the matter fields are introduced as fields
transforming as modular functions of a given weight and show that their
$(mass)^2$  is negative signalling that the vacuum is unstable, as
long as their modular weight belongs to a given range. The twisted
matter fields in orbifold models have weights in such a range, so
vacua are unstable.  In section four, the existence of stringy cosmic
domain walls is studied \CQRI.  First the example of global supersymmetry
with duality invariant potential is
discussed and then the full $N=1$ supergravity case is discussed.
Conclusions are given in section five.

\chap{Duality Invariant Lagrangian Effective Supergravity}

\noindent\undertext{\caps Target Space Duality}

In this section we shall restrict ourselves to a single modulus field
$T\equiv R^2+iB$ which respects $SL(2,{\bf Z})$ duality symmetry.  The
radius $R$ corresponds to the radius of the complex six-dimensional
compactified space and $B$ is the internal axion field.  The K\"ahler
potential has the form:
$$K(T,T^\ast)=-3\log(T+T^\ast)\eqn\twopone$$
while there is no superpotential for the $T$ fields
$$W\equiv0.\eqn\twoptwo$$
Effectively the space of moduli field $T$ is described by the coset
$SU(1,1)/U(1)$ and the quantum duality group is given by $SL(2,{\bf
Z})$. This group acts on the complex modulus $T$ is defined as the group
of modular transformations
$$\Gamma: T\rightarrow T^\prime={{aT-ib}\over{icT+a}} \quad
a,b,c,d\in {\bf Z},\quad
ad-bc=1.\eqn\twopthree$$
Actually the relevant group is $PSL(2,{\bf Z})=SL(2,{\bf Z})/{\bf Z}_2$.
$\Gamma$ acts as a K\"ahler transformation on $K$:
$$K\rightarrow K^\prime=K+f(T)+f^\ast(T^\ast),
f=\log(icT+d)^3.\eqn\twopfour$$
On the other had the $T$ dependent superpotential can only be due to
nonperturbative (infinite genus) physics. We will now use the discrete
quantum symmetry \twopthree\ to strongly constrain the
the possible form of $W(T)$ regardless of what actual non-perturbative
mechanism is responsible for the emergence of the superpotential.
For non-vanishing superpotential the $N=1$ supergravity action is
described \CREMMER\ by the function
$$G(T,T^*)=K(T,T^*)+\log |W(T)|^2.\eqn\gfct$$
We also assume that the string tree-level transformation behaviour
Eq.~\twopfour\ of $K(T,T^*)$ is not changed, including the non-perturbative
contribution to the superpotential.
This assumption is well justified
in the weak coupling limit with small but non-vanishing $W(T)$.
Then the duality invariance of the effective action requires $G(T,T^*)$
to be duality invariant, as well, thus
implying that $W(T)$ must transform under
$\Gamma$ up to a phase as
$$W(T)\to W(T) e^{-f(T)}.\eqn\suppottrans
$$
This means that $W(T)$ is related to the automorphic functions of the
discrete duality group $\Gamma$.

For $\Gamma=SL(2,{\bf Z})$, the superpotential has to transform
under target space modular transformations as
$$W(T)\to e^{i\delta (a,b,c,d)}{W(T)\over (icT+d)^3}.\eqn\mod
$$
Thus, $W(T)$ has to be a modular function of
modular weight $-3$. The field independent phase $\delta (a,b,c,d)$
is the so-called multiplier system.
$W(T)$ can be written in the following form:
$$W(T)= { H(T)\over \eta(T)^6 }.\eqn\wht
$$
Here $\eta(T)$ is the Dedekind function and $H(T)$,
being an arbitrary modular invariant function with in general
non-vanishing multiplier system, is a rational function
of the absolute modular invariant $j(T)$.
Note that Eq.~\wht\ implies that $W(T)$ has an expansion
in $e^{T/\alpha '}$ (for large $T$) after reintroducing the dimensionful
slope parameter $\alpha '$. This form is obviously
non-perturbative in the two-dimensional
$\sigma$-model coupling constant $\alpha '/
R^2$ and clearly shows the relevance of the contribution
of the world-sheet instantons to $W(T)$.
\par
The behaviour of $W(T)$ will largely determine the vacuum structure
of the $N=1$ supergravity action. In particular, the scalar potential
will inherit its singularities.
In fact,
since  $W(T)$ has negative modular weight it follows that
$W(T)$ will in general have singularities at $\infty$ and/or
at finite values of $T$.

\topinsert
\vskip9cm
\setbox1=\vbox{
\hsize=12cm\raggedright
\noindent{\bf Fig.1.} The standard fundamental domain ${\cal D}$
for the $T$-field. The boundary of the moduli
space of the $PSL(2,{\bf Z})\times{\bf Z}_2$ transformations
is denoted by ${\cal B}$ where all minima of the scalar potential
are expected to lie.}
\hfill\box1\hfill
\medskip
\hrule
\endinsert

According to a mathematical theorem quoted in \LEHNER,
if we want $W(T)$ to avoid
singularities inside the fundamental domain we must take $H(T)$ of the
form
$$
H(T)=
 \left ( {G_6(T)\over {\eta(T)^{12}}} \right )^m
 \left ( {G_4(T)\over {\eta(T)^8}} \right )^n {\cal P}(j)
 \eqn\hta
$$
or equivalently (up to a constant) as
$$ H(T) = (j-1728)^{m/2} j^{n/3} {\cal P}(j).\eqn\htb
$$
Here $m,n$ are positive numbers and ${\cal P}(j)$ is a polynomial of
$j(T)$.
$G_4(T)$ and $G_6(T)$ are the Eisenstein functions of
modular weight 4 and 6 respectively (see Ref.~\LEHNER\ for
details).
Without loss of generality the zeroes of ${\cal P}(j)$ are distinct from
the fixed points $T=1, T=\rho \equiv e^{i\pi/6}$. This means that
$H$ vanishes at $T=\rho$ only if $n\geq 1$ and at
$T=1$ only if $m\geq 1$. Notice that superpotentials of this type
always diverge at the cusp point $T \to \infty$ and at the
modular transformed point $T=0$ (see Fig.~1 for the fundamental domain
${\cal D}$ of $T$).
\par
So far we have only discussed the dependence of the superpotential
on the internal
$T$-field modulus. To discuss the issue  of supersymmetry breaking
in string theory one has also to include
the $S$-field, $S={1\o g^2}+i\theta$,
whose $vev$
determines the string coupling constant $g$ and the space-time
axion field $\theta$. We take for the $S$-field K\"ahler potential its
tree-level form which is at least justified for small couplings:\foot{It
is conceivable that one can redefine the $S$-field in any order
of string perturbation theory such that the K\"ahler potential
always takes the tree-level form. In this approach we neglect the mixing
between $S$ and $T$ fields which is due to the Green-Schwartz-type
mechanism studied in \multref\DFKZ{\CO}. See also \LOU.}
$$K(S,S^*,T,T^*)=-\log(S+S^*)-3\log(T+T^*).\eqn\kpst
$$
The non-perturbative effects involving the $S$-field are parametrized
by the superpotential:
$$W(S,T)={\Omega(S)H(T)\o \eta(T)^6}.\eqn\supst
$$
Here $\Omega(S)$ is a function of the $S$-field.
For the sake of simplicity we assume here that the dependence on the
dilaton field $S$ factorizes.
Note however since the superpotential \supst\ is due to
non-perturbative string effects, $\Omega(S)$ should have an
expansion in terms of powers of $e^{-S}$.
\par
The scalar potential resulting from eqs.\kpst,\supst\ is given by
$$\eqalign{V&=|h^S|^2G_{SS^*}^{-1}+|h^T|^2G_{TT^*}^{-1}-3\exp(G)\cr
&= {1\over {S_R T_R^3 \vert \eta\vert ^{12} } } \left \{
\vert S_R\Omega_S - \Omega\vert ^2 \vert H\vert ^2
 + {T_R^2\over 3} \vert H'
 + {3\over {2\pi}}H \hat G_2\vert ^2 \vert \Omega\vert ^2
  - 3\vert H\vert ^2\vert \Omega\vert ^2 \right \},}
   \eqn\scalpot
$$
where $T_R=T+T^*$, $S_R=S+S^*$,
$H' = dH/dT$, $\Omega_S=d\Omega /dS$ and $\hat G_2\equiv-4\pi{{\partial
\eta}\over{\partial T}}{1\over\eta}-{{2\pi}\over T_R}
$ is the
non-holomorphic Eisenstein function of modular weight two.
Space-time supersymmetry is spontaneously broken if at the minimum
of $V$ one of the auxiliary fields
$$h^i=\exp({1\o 2}G)G^i
=|W|\exp({1\o 2}K)\Biggl( K^i+{W^i\o W}\Biggr)\eqn\aux
$$
has a non-zero vacuum expectation value.
\par
\noindent\undertext{\caps Properties of Duality Invariant Potentials}

Now let us concentrate on the vacuum structure with
respect to the $T$-field.
For $m,n\geq 0$ and ${\cal P}(j)$ a polynomial the only singularity of
$V(T,T^*)$ is a pole at the cusp point $T\to \infty$ .
At this point
the potential goes to $+\infty$ since for large $T$ the term
$|h^T|^2G_{TT^*}^{-1}$ dominates over $-3|W|^2$
independently of the form of $H(T)$. Since $V$ is modular invariant
it must also have a pole at $T=0$.
Therefore we obtain the model-independent result that for
this class of regular
superpotentials the non-perturbative dynamics always fixes
the $vev$ of the modulus $T$. The theory is forced to
be compactified since the minimum of $V$ always occurs at
finite radius. Furthermore, generically $V < 0$ at the global
minimum since the potential is not positive definite.
\par
Another generic property of $V$ is the fact that the modular
fixed points $T=1,\rho$ must be extrema. This is because
$\partial V / \partial T $ transforms with modular weight two and
has no singularities inside the fundamental domain.
The self-dual points $T=1,\rho$ are
extrema but not necessarily local minima
of $V$. To develop a better understanding of the shape of $V$ it is
useful to study its behaviour at the self-dual points.
Let us consider $T=\rho$ first. When $n \geq 2$, $T=\rho$ is a
minimum with $V=0$.
When $n=0$ it is a minimum if
$Z(S) \equiv  \vert S_R\Omega_S - \Omega\vert ^2
 - 2\vert \Omega\vert ^2 > 0$ and a maximum if
$Z(S) < 0$. On the other hand for $n\leq1$, $T=\rho$, \ie\ supersymmetry
is generically preserved at saddle points.
We now turn to $T=1$.
When $m \geq 2$ it is a
minimum with $V=0$ and $h^T=0$.
While for $n\leq1$ this point is generically a saddle point.

In all the above cases, the minima found at $T=\rho , 1$ have
unbroken SUSY (at least in the $T$-sector). Notice that,
although the minima are supersymmetric, the value of the scale of
compactification is determined. This shows that target-space modular
invariance allows for a situation in which non-perturbative effects
lift the degeneracy of the compactification radius but do not break
supersymmetry.

\topinsert
\vskip11.cm
 \setbox2=\vbox{\hsize=6.5cm\raggedright
\noindent{\bf Fig.2.}
The scalar potential $V$ as a function of $({\rm Re}T,{\rm Im}T)$
for ${\cal P}=1$ and $(m,n)=(0,0)$. The negative energy minimum
lies on the real
axis at $T\sim 1.2$ (and at the modular transformed images).}
\setbox3=\vbox{\hsize=6.5cm
\noindent{\bf Fig.3.}
The scalar potential $V$ as a function of $({\rm Re}T,{\rm Im}T)$
for ${\cal P}=1$ and $(m,n)=(2,2)$. The zero energy
minima are $T=1$ and $T=\rho$.}
\hbox{\box2\hbox to0.8cm{ }\box3}
\medskip
\hrule
\endinsert

We have studied numerically several cases with ${\cal P}=1$ and various
values of $m,n$. In these examples we also assumed
$(S_R\Omega_S - \Omega)=0$. In all cases we verified that $V < 0$ at
the global minimum which occurs at points on the boundary ${\cal B}$.
Examples with ${\cal P}=1$ and $(m,n)=(0,0),(2,2)$
are shown in figures 2,3.
For $(m,n)=(0,0)$, the case already shown in \FILQ, the minimum
lies on the real axis, $T_{\rm min}\sim 1.2$.
For $(m,n)=(2,2)$ there
are two minima with unbroken supersymmetry. They are both degenerate
with zero energy and both are located at the self-dual points,
$T_{\rm min}=1$ and $T_{\rm min}=\rho$.

\noindent\undertext{\caps The Vanishing of the Cosmological Constant}

As we have seen in the last section, for general choices of the
modular invariant function $H(T)$, including $H(T)={\rm const}$,
the scalar potential takes a negative value at its supersymmetry
breaking minimum. Thus we are dealing in general with a negative
cosmological constant after supersymmetry breaking; four-dimensional
space-time is described by an anti-de Sitter space. (From the conformal
field theory point of view the negative cosmological constant can
be regarded as a shift of the central charge $\delta c<0$, such
that we are dealing with a non-critical string after taking into account
the non-perturbative effects.) Obviously, it is a very interesting
question whether there exist special choices for $H(T)$ with broken
space-time supersymmetry and vanishing scalar potential at some
(possibly local) minimum $T=P$.

A class of modular invariant functions $H(T)$ with these
properties can be easily constructed by a third-order polynomial
in $j(T)$:
$$H(T)=j(T)^3+aj(T)^2+bj(T)+c.\eqn\solution
$$
The coefficients $a,b,c$ are uniquely determined by a system
of three linear equations.

Thus, one can find a special class of superpotentials
which leads to a supersymmetry breaking minimum of the corresponding
scalar potential with vanishing cosmological constant.
In order to find this minimum the auxiliary field
$h^S$  must be non-zero.
Apart from this zero energy minimum \solution\
leads to additional minima with broken supersymmetry and
negative cosmological constant.

\noindent\undertext{\caps Different Radii for Two-Tori}

Now one would like to study  the case with the three-moduli $T_i$
$(i=1,2,3)$ corresponding to the radii of the three two-tori of the
internal space.  The connection with the overall radius $T$ is the
following:
$$(T+T^*)^3=\prod_{i=1}^3(T_i+T_i^*).\eqn\deft$$
One can show that
when the choice for the superpotential is
$W=\Omega(s)\prod_{i=1}^3\eta^{-2}(T_i)$,
the preferred minimum occurs when $T_1=T_2=T_3=T$.
This in turn gives support to studying only one modulus $T$.
Although  such a conclusion seems to be
obvious starting from a Lagrangian which is completely symmetric
under the exchange of the three $T_i$,
there are examples of spontaneous breaking of  similar
discrete  symmetries within the context of gauge theories beyond the
standard model. An exemplary case is the spontaneous breaking of the
left-right gauge symmetry \multref\PASA{\MOSEN}. It is thus important
that one has verified that such a spontaneous symmetry breaking
does not happen in our case.

\chap{Matter Fields}

So far we have been consistently neglecting the matter fields,
$i.e.$ we have been setting them to the naive vacuum state where
all of them vanish. In orbifold compactifications, these fields come
in three classes: the untwisted matter ($A$) corresponding   to the Wilson
lines of toroidal compactifications which survive the orbifold
twist; some of them are actually moduli since they may have flat
potential also;
the twisted moduli or blowing-up modes ($B$) present on $(2,2)$
(and some $(0,2)$)
models, whose $vev$ determines if we are at the orbifold point ($B=0$)
or at any other smooth manifold point, and finally we have the twisted
matter fields ($C$), which like untwisted ones may or may not
have a flat potential. They have been the ones responsible for breaking
an original gauge group $G$ to the Standard Model one in the
quasi-realistic orbifold models constructed so far \FINQ.

Since all of these fields are massless to start with, the original
(perturbative) potential is such that their mass terms in the scalar
potential vanish. It is then interesting to investigate if the
non-perturbative induced potentials change this situation by providing a
$(mass)^2$ to these fields which, if it is negative, will
prove that the original vacuum was a maximum in these directions.
It takes the following form \multref\DKL{\FLT\CLOA\CMO\CVETIC}:
\foot{This equation applies
if all three $U(1)^3$ charges (see ref.\DKL ) are
non-vanishing. If this is not the case, there are further twisted
moduli, besides $B$, which for our purposes will behave like the
field $A$ above so we do not consider them explicitly.}
$$ K= -\log\{ (T+T^*-AA^*)^3-BB^*-(T+T^*)CC^*\}.\eqn\kpmatt
$$
With respect to the twisted fields $B$ and $C$, eqn.\kpmatt\ provides
only the first term in a perturbative
expansion of $K$
around zero.\foot{For the twisted moduli $B$ higher order corrections
to $e^{-K}$ were discussed in ref.\FLTA.}
This however will be enough for our purposes
in this chapter since we will only check if the naive vacuum is stable
under small perturbations. What we will not be able to find is some
other extrema for larger values of the fields.
It is known that all of the matter fields transform
non-trivially under the
$SL(2,{\bf Z})$ modular group. In particular the untwisted matter
field transformations can be found directly from the generalized
torus duality of references \GRV,\SHWI.
We can see that they transform
as modular forms of weight $1$ (up to a phase):
$$ A\rightarrow (icT+d)^{-1} A \eqn\atransf
$$
For the twisted fields $B$ and $C$, we can find their behaviour
under modular transformations by demanding the K\"ahler potential
to transform adequately \FLT :
$$\eqalign{B\rightarrow (icT+d)^{-3} B, \cr
           C\rightarrow (icT+d)^{-2} C. \cr} \eqn\bctransf
           $$
up to a $T$-dependent phase (multiplier system). Therefore
$A$, $B$ and $C$ transform like modular forms of weight -1, -3
and -2 respectively. For simplicity, we
will consider a generic field $Z$ transforming as a modular form of
weight $-n$ and expand the potential around $\langle Z \rangle =0$.
We will then consider the following K\"ahler potential:
$$K(S,S^*,T,T^*,Z,Z^*)=-\log(S+S^*) + K_0 (T,T^*) +
K_1(T,T^*)ZZ^*+\cdots .\eqn\kppertu
$$
Notice from \kpmatt\ that for
$A$, $B$ and $C$ we have
$K_0=-3\log(T+T^*)$ and $K_1=\alpha_n (T+T^*)^{-n}$, $\alpha_n \geq 0$
which is consistent with the positivity of the kinetic energy terms.
Our aim is then to evaluate the scalar potential $V$ in a neighbourhood
of $Z=0$
$$V=V_0(S,S^*,T,T^*) + V_1(S,S^*,T,T^*)ZZ^*+\cdots \eqn\scalpotmatt
$$
and find the explicit signature of $V_1$ in order to determine
the stability of the vacuum.
{\it E.g.,} $n=1,2,3$, for untwisted matter fields (A), twisted matter
 fields (C) and
blowing-up modes (B), respectively.
Then terms in eqs. \scalpotmatt\  become:
$$\eqalign{  V_0=& {1\over{(S+S^\ast)(T+T^\ast)^3}}
   \left\{ (S+S^\ast)^2 \vert
    D_S\Omega\vert^2\vert W_0\vert^2+ {{(T+T^\ast)^3}\over3} \vert
    D_T W_0\vert^2\vert\Omega\vert^2\right.\cr
&\left.\phantom{ {1\over{(S+S^\ast)(T+T^\ast)^3}}\{}
   -3\vert W_0\Omega\vert^2 \right\} ,\cr
V_1=& {{\alpha_n}\over{(T+T^\ast)^{n+3}}} {1\over{(S+S^\ast)}}
    \left\{ \vphantom{ {(T+T^\ast)^2\over3} }
   (S+S^\ast)^2\vert W_0\vert^2\vert D_S\Omega\vert^2+ \right.\cr
&\left.
   \phantom{ {{A_n}\over{(T+T^\ast)^{n+3}}} {1\over{(S+S^\ast)}} \{}
   {(T+T^\ast)^2\over3}
   \vert\Omega\vert^2\vert D_T W_0\vert^2
   \left(  1-{n\over3} \right) -
    2\vert\Omega W_0\vert^2 \right\}.\cr} \eqn\vpertu
$$
For the minimum with $D_S\Omega\equiv0$, which has been shown in the
previous section to be the absolute
minimum for $\Omega =c+ h\ e^{(-\alpha S)}$
one sees from eqs.\vpertu\ that
as long as  $n\ge1$, $V_1<0$\rlap.\foot{Note, that for $n=1,2$
we took into account $V_0<0$.} Thus, $m^2$  terms
are negative for all the three types of the
``matter'' fields.  This result is
{\it independent} of the explicit form for the non-perturbative
supersymmetric potential $W_0$ and it depends
only on the structure of the
K\"ahler potential. One can also check that for $V_0\rightarrow 0$
(limit of vanishing cosmological constant) the untwisted
matter fields become massless.

There is a subtle issue  associated with the
stability of  the anti-de Sitter vacuum, relevant for most of our cases
with $V_0<0$. Namely, the negative $m^2$ does not
automatically imply that the vacuum is not stable.
In flat space-time, the criterion for stability of a
static field configuration is well known to be parametrized
by having positive $m^2$. In curved backgrounds, the situation is
less clear since there are gravitational  corrections to the energy.
For the anti-De Sitter
space the condition for the stability of the vacuum has been discussed
in the past \BRFR , and the criterion is the following. For a massive
field $\Phi$ in $D$ dimensions  satisfying
$$ ( \square^2+ M^2 )\Phi = 0, \eqn\kg$$
the configuration is stable if
$$ M^2 > -{1\over4}(D-1)^2 \kappa^2 \eqn\stab
$$
where $\kappa$ is the constant determining the curvature of the space
$R_{\mu\nu\lambda\rho}=\kappa^2 \left(g_{\mu\rho}g_{\nu\lambda}\right.$
$\left.-  g_{\nu\rho}g_{\mu\lambda} \right) $. From Einstein's equations
we can see that the cosmological constant $\Lambda \equiv V_0 =
-{\kappa^2\over 2} (D-1)(D-2)$. In four dimensions $\Lambda=-3\kappa^2$,
 so the stability criterion can be expressed as
$$ M^2 > {3\over 4}\Lambda .\eqn\staba
$$
In the present case, we have a Lagrangian
$$\alpha T_R^{-n}\partial Z\partial Z^* + V_1ZZ^* \eqn\langra
$$
so $M^2={V_1\over{\alpha T_R^{-n}}}$. For a field of modular weight
$-n=-1$, we know that $V_1 = {2\over3}\alpha T_R^{-n} V_0$ so
$M^2= {2\over3}V_0 > {3\over4}V_0$ and this maximum is then
stable. For more negative weights however the $V_1$ term is
``more negative'' and they will be generically unstable (as can
be easily checked for particular cases on the blowing-up modes
and twisted matter fields).

We have seen then that the above  constraint is barely not
satisfied for the
untwisted matter fields (A), however it is possible  to satisfy
it for the  twisted matter (C) and the blowing-up modes (B).
Actually, for the realistic case when the cosmological
constant is extremely small , {\it i.e.} $V\rightarrow0$,
B and C  fields have strictly negative $m^2$, while the
mass of A fields goes to zero. Thus our
conclusion that the blowing-up modes and the twisted matter fields
destabilize the vacuum is generic.

\chap{Stringy Domain Walls}

Topological defects are known to be present in phase transitions
triggered by spontaneously broken symmetries of a given theory. They
can have very important physical implications, especially for
cosmological consequences of the theory. It is thus interesting to
explore the existence of these configurations in four-dimensional
string vacua.
Recently, cosmic string solutions were found in perturbative string
theory \vafastring. We will present domain wall solutions
from nonperturbative potentials in four-dimensional string vacua
\CQRI.\foot{Domain walls in string theory have been studied previously
in the context of broken discrete symmetries in Calabi-Yau models in
\CHOI.}

In fact, the physics of moduli fields
is an intriguing generalization of the well-known axion physics\axion\
 introduced to solve the strong CP problem in QCD.
Below a scale $f_a$, the $U(1)$ Peccei-Quinn symmetry is spontaneously
broken  and it
is realized nonlinearly through a pseudo-Goldstone boson,
the invisible axion $\Theta$.
 The low-energy effective Lagrangian of the invisible
 axion is described by
 $$
 L_{\rm eff}
 = {f_a^2 \over 2} (\nabla \Theta)^2 + {1 \over 4 g^2} F_{\mu \nu}^a
 F^{\mu \nu a} + {N_f \over 32 \pi^2} \Theta  F_{\mu \nu}^a
 \tilde F^{\mu \nu a}.
 \eqn\fourpone
 $$
The non-linearly realized Peccei-Quinn $U(1)$ symmetry,
$\Theta \rightarrow \Theta + \alpha$, where $\alpha=constant$, is
\sl explicitly \rm broken due to the nonperturbative QCD effects through the
axial anomaly $\partial_\mu J^\mu_5 = {N_f \over 32 \pi^2} F_{\mu \nu}^a
\tilde F^{\mu \nu a}$.
Integrating out the instanton effects by a dilute instanton approximation
\multref\original{\ovrut}\
generates an effective potential proportional to $ \ \ 1-\cos N_f\Theta\ \ $
 with  degenerate  minima at
$\Theta = {2 \pi k \over N_f}$, where $k = 0, 1, \cdots, N_f - 1$,
thus breaking the original $U(1)$ Peccei-Quinn symmetry  down to a
\sl discrete \rm subgroup $Z_{N_f} \in U(1)$.
 Note that, within the dilute instanton approximation, the
form of nonperturbatively generated axion potential is completely determined
by the invariance  under the residual discrete subgroup
$Z_{N_f}$. It is well known that these potentials lead to domain wall
solutions  between adjacent vacua\vilen.
$N_f$ domain walls meet at the axionic string.

In our case,
the modulus field  $T$ possesses a non-linearly realized
noncompact symmetry $SU(1,1)/U(1) \approx SL(2, {\bf R})/U(1)$
to all orders in the sigma model perturbation.
The real and the imaginary part of the modulus field $T$ are nothing
 but the
Goldstone bosons associated with spontaneously broken dilatation and
axial symmetries, which are generalizations of the Peccei-Quinn symmetry
in the invisible axion physics. In a completely analogous way the
world-sheet instanton effects\worldsheetinstanton\
or space-time
axionic instanton effects break the above
non-linearly realized global symmetry to a discrete subgroup of it,
$PSL(2, {\bf Z})$. Therefore, the superpotential should be a
holomorphic function of $T$ transforming
covariantly  under the above discrete subgroup. Indeed, it is known
that the modular invariance puts  strong constraints on
the dynamical supersymmetry breaking and the stability of the
 vacua \bigpaper.

   As an instructive and illuminating example we
first consider a global supersymmetric theory by turning off
 gravity.
The effective Lagrangian reads
$$
\eqalign { L &= \int d^2 \theta d^2 \bar \theta K(T, \bar T)
+ \int d^2 \theta W(T) + \int d^2 \bar \theta \bar W( \bar T) \cr
& = G_{T \bar T} |\nabla T|^2 + G^{T \bar T} |\partial_T W(T)|^2}
\eqn\oldtwo
$$
Here,
$ G_{T \bar T} \equiv \partial_T \partial_{\bar T} K(T, \bar T)\  $
is the positive definite metric on the complex modulus space and
$W$ the superpotential,
which has to be  a modular invariant
(weight zero) form of $PSL(2,\bf Z)$
 defined over the  fundamental domain
$\cal D$ of the  $T$-field.
The most general form of the superpotential
is a rational polynomial $ P(j(T))$
of the modular-invariant
function $j(T)$\modularform.

Since the symmetry group is a discrete group\giveon\
 it is natural to
expect that the semi-positive definite potential
$$
 V\equiv G^{T \bar T} |\partial_T W(T)|^2 = G^{T \bar T}
|\partial_j  P(j) \partial_T j(T)|^2
\eqn\oldthree $$
 has a discrete set of degenerate minima,
and thus there is a stable domain wall solution
\foot{In particular, the  symmetry
$V(T+in)=V(T)$ with $n\in Z$ ensures that for each minimum at $T_0$
in the fundamental domain there is a degenerate one for $T_0+in$. However,
as we will see later, it is crucial for the nontrivial domain wall
solution that the nearby minima are at the points $T$
which give  different values for the superpotential.},
i.e. in this case the homotopy group $\Pi_0 (\cal M)$ is non-trivial.
The term $|\partial_T j(T)|^2$ has
two isolated zeros
at $T = 1$ and $T=\rho\equiv e^{i \pi/6}$
in the fundamental domain $\cal D$ for $T$ \modularform.
Other isolated degenerate minima might as well arise when
$|\partial_j   P(j)|^2=0$.

We embed the domain wall in the  $(x, y)$ plane,
which is thus perpendicular to the $z$ direction.
Then, the mass per unit area of the domain wall is
$$
\mu \equiv { E \over \int dx dy}=\int_{-\infty}^\infty \!
dz \,\, [ G_{T \bar T} |\partial_z T|^2 + G^{T \bar T} | \partial_T W(T)|^2]
\eqn\oldfour$$
We now look for a static, spatially nontrivial field configurations that
minimize the mass per unit area in Eq. \oldtwo.
We can rewrite Eq.\oldfour\
as\rlap:\foot{It is intriguing that the present kink solitons also
appear in integrable, supersymmetric two-dimensional $N=2$
Landau-Ginzburg models \vafa.}
$$
\mu = \int_{-\infty}^{\infty} dz\, G_{T \bar T}
| \partial_z  T - e^{i\theta} G^{T \bar T} \partial_{\bar T}
\bar W(\bar T)|^2  + 2 Re (e^{- i \theta} \Delta W)
\eqn\oldfive
$$
where
$\Delta W \equiv  W(T(z = \infty)) - W(T(z= -\infty))$. The arbitrary
phase
$\theta$ has to  be chosen such that
 $e^{i \theta} = \Delta W / |\Delta W|$,
thus maximizing the cross term in Eq. \oldfive.
Then, we find $\mu \ge K \equiv 2 |\Delta W|$, where $K$ denotes the kink
number. Since $\partial_T W$
is analytic in $T$, the line integral over $T$ is \sl path independent \rm
as for a conservative force. The minimum is obtained only if
the Bogomolnyi bound
$$
\partial_z T(z) = G^{T \bar T} e^{i \theta} \partial_{\bar T} \bar
W(\bar T(z))
\eqn\oldsiz
$$
is saturated. In this case
$
\partial_z W(T(z)) = G^{T \bar T} e^{i \theta} |\partial_T W(T(z))|^2
$,
which implies that the phase of $\partial_z W$ does not change with $z$.
Thus, the supersymmetric domain wall is a mapping from the
 z-axis $[-\infty,
\infty]$ to a \sl straight line \rm  connecting between two degenerate
vacua in the $W$-plane. The domain wall is stabilized by the topological
kink number $K =   \pm 2\times |\Delta W|$.
We would like to emphasize that this result is general; it
applies to any globally supersymmetric theory with disconnected
degenerate minima that preserve supersymmetry.

The above observation is  neatly described
by using the supersymmetry
transformations of the moduli-superfields. The moduli
field Lagrangian possesses an enhanced $N=2$ spacetime
supersymmetry \DKL. The nonrenormalization
theorem of $N=2$ supersymmetry in turn
guarantees no quantum correction to the
mass density of the stringy domain walls\rlap.\foot{The very same
argument has been used earlier in the case of axionic instantons \WO.}
This can be seen as follows. Denoting the `modulino' as $\chi$,
supersymmetry charges are
 $$
 {\bf Q}_\alpha = (\gamma^\mu \chi)_\alpha \nabla_\mu T + \chi_\alpha
 G^{T \bar T} \partial_{\bar T} \bar W(\bar T),\ \
 \bar {\bf Q}_\alpha = (\bar \chi \gamma^\mu)_\alpha \nabla_\mu \bar T
 + \bar \chi_\alpha G^{T \bar T} \partial_T W(T).
\eqn\oldseven$$
The anticommutator of supercurrents contains the Hamiltonian density
and a total derivative, central charge term. The latter is nothing but
the aforementioned kink number. We now introduce a constant, chiral spinors
$\epsilon_\pm$ of unit norm  $\bar \epsilon_\pm \cdot \epsilon_\pm  = 1$.
Then, in the rest frame of domain wall,
 we find
$$
\mu \mp K = {1 \over 2} \int dz \, \bar \epsilon_\pm \{ {\bf Q}_\alpha,
\bar {\bf Q}_\alpha \} \epsilon_\pm \ge 0.
\eqn\oldeight
$$
Here, $K$ denotes the kink number we derived in Eq.\oldfive.
Thus, the Bogomolnyi bound is achieved for $\delta_\epsilon \chi
\equiv \epsilon \cdot
{\bf Q}\, \chi = 0$. It is identical to Eq. (6) in which the $\theta$ is
identified with the relative phase between $\epsilon_+$ and $\epsilon_-$.

We shall now illustrate the above points with the explicit choice for
the superpotential
$W (T) = (\alpha')^{-3/2} j(T)$ with $\alpha'$ being the string tension.
As already discussed before, in the fundamental domain $\cal D$
 the potential has two isolated  degenerate
minima at $T = 1$ and $T=\rho\equiv e^{i \pi/6}$.
 At these fixed points,
$j(T=\rho) = 0$ and $j(T=1) = 1728$. Therefore, the mass per unit area
is $\mu = 2 \times 1728 (\alpha')^{-3/2}$.
Other cases can be worked out analogously.

Naive application of Eq. \oldfive\ implies that the domain
wall solution between  the minima
that are connected by the $PSL(2,\bf Z)$ transformations
has zero  energy stored
since $W$ has the same value at those points. However, one
can show  that in the fundamental domain $\cal D$ (see Fig.~1)
there are always {\it at least} two degenerate minima with
different value of the superpotential, and thus the
energy density
of the domain wall that interpolates between these two
minima is {\it nonzero}. The energy density of the domain walls
interpolating between the minima connected by the $PSL(2,\bf{Z}) $
transformations are  thus in turn determined by taking the path
 through all the minima in between. This adjusts
the constant phase $\theta$
between the  adjacent minima to maximize
 the cross-term in Eq. \oldfive.
For example, for
$W (T) = (\alpha')^{-3/2} j(T)$
the energy density stored   in the domain wall
that interpolates between $T=e^{i\pi /6}$ and $T=e^{-i\pi/6}$
is $\mu =2\times 2\times 1728 (\alpha')^{-3/2}$.

   We shall now study the case with  gravity restored.
The structure of $N=1$ supergravity action was discussed in section two.
There it was observed that for superpotential  with
$m\geq 2,\  n\geq 2$ and $ P(j)=1$, the potential is (see Fig.~3)
semi-positive
definite with the two isolated minima at $T=1$ and $T=\rho$
and unbroken local   supersymmetry just like in the global
supersymmetric case.

 We now minimize the domain wall mass density. By the planar
 symmetry, the most general {\it static}
 Ansatz  for  the metric \domainansatz\   is
 $ ds^2 = A(|z|) (-dt^2 + d z^2) + B(|z|) (dx^2 + dy^2)$
 in which the domain wall is oriented parallel to $(x,y)$ plane.
Using the supersymmetry transformation laws
$$
\eqalign{\delta \psi_{\mu\alpha}
& = [\nabla_\mu(\omega)  -{ i \over 2} Im
(G_T \nabla_\mu T)] \epsilon_{\alpha}  + {1 \over 2}
(\sigma_\mu \bar \epsilon)_\alpha  e^{G \over 2}\cr
\delta \chi_\alpha  &= {1 \over 2}(\sigma^\mu \bar \epsilon)_\alpha
\nabla_\mu T
 - e^{G \over 2}
G^{T \bar T} G_{\bar T} \epsilon_\alpha}
\eqn\oldthirteen
$$
with commuting, covariantly constant, chiral spinors $\epsilon_\pm$,
the ADM mass density $\mu$ can be expressed as \positiveenergy:
$$
\mu \mp K =  \int \! dz \, \sqrt g [ g_{ij}
\delta \psi^{\dagger i}\delta \psi^j + {1\over 2} G_T\bar T
\delta \chi^\dagger \delta \chi] \ge 0.
\eqn\oldfourteen
$$
The $i,j$ indices are for spatial directions.
The minimum of the Bogomolnyi bound is achieved if Eq.~\oldfourteen\
 vanish.
Again, the stringy domain wall is stabilized by the topological kink number.

Unfortunately, the nice holomorphic structure of the scalar potential is
lost. In other words, there is now
 a \sl holomorphic anomaly \rm in the
scalar potential due to the supergravity coupling.
This implies that the path connecting two
degenerate vacua in superpotential space is \sl not \rm  a straight line.
In fact, one can understand the motion as a {\it geodesic}
 path in a nontrivial
K\"ahler metric, thus in $G(T, \bar T)$.
One can show (numerically) that
in  our example the path along the circle $T= \exp{i\theta(z)}$ with
$\theta= (0, \pi/6)$, $i. e.$ , the
 self-dual line of $T \rightarrow 1/T$ modular transformation, is the \sl
geodesic \rm
path connecting between $T=1$ and $T=\rho$ in the scalar potential space.
Thus, we have again established an existence of stable domain walls.
The superpotential
is quite complicated, however the numerical solution can be
obtained\ours.

It is interesting to note that  stringy cosmic strings \vafastring\
can be viewed as boundaries of our domain walls. Because the domain wall
number is two, the intersection of two such domain walls is precisely
the  line of stringy cosmic strings. On the other hand such stable
domain walls are disastrous from the cosmological point of view.
One possible solution to this problem  is that after supersymmetry
breaking, the degeneracy of the two minima is lifted. In that case,
the domain wall becomes unstable via the false vacuum decay\falsevacuum.

\chap{Conclusions}

We have seen how powerful the constraint of $SL(2,{\bf Z})$
modular invariance is for the low-energy string effective actions.
The most general invariant superpotential that we considered
still has a large amount of degrees of freedom which parametrize
our ignorance of the sources of non-perturbative string effects.
Nevertheless, we have been able to extract general properties
which will hold for any modular invariant string action.
Probably the most striking one is that due to the fact that the
superpotential is a modular form of negative weight, then it
has to have singularities either inside the fundamental domain or
at infinity.

When the scalar field was massless
but with non-flat potential, inducing a negative $(mass)^2$ will
indicate that the global minimum corresponds to a value of the field
which was not classically allowed, {\it{i.e.}} it did not
correspond to a conformal field theory.
For $(2,2)$ orbifold models,
the blowing-up modes acquire          a
non-vanishing $vev$, thus
leading
 to a nice geometrical interpretation
where a smooth Calabi-Yau point is preferred over the
singular orbifold. Notice that unlike the untwisted fields a
negative $(mass)^2$ for the twisted modes is guaranteed
independent of the value of the cosmological constant. That
is, even in the cases with vanishing vacuum energy, the
potential has a relative maximum in those directions.

As for the breakdown of supersymmetry we have seen that it
is very generic and far from trivial that the global minimum breaks
supersymmetry. The scale of breaking is
actually unknown since it depends mostly on
dilaton dependence of the superpotential which  we cannot restrict.
There is also
enough freedom in
the superpotential to have a non-supersymmetric minimum with
vanishing cosmological constant which is very encouraging. On the
other hand there is no criterion based on modular invariance alone
to prefer those superpotentials to the ones leading to
anti-de Sitter space.

Another intriguing application of our analysis is the observation that
for a class of duality invariant potentials
there are two degenerate supersymmetric minima at $T=1$
and $T=\rho$ which allow for the stable supersymmetric
stringy domain wall solution
with interesting cosmological implications.

\vskip1.cm
{\bf Acknowledgements:}
I would like to thank my collaborators A.~Font, S.~Grif--
fies,
L.~Iba\~nez, D.~L\"ust,
F.~Quevedo, and S.-J.~Rey for an enjoyable collaboration and numerous
discussions.
Research was supported in part by the Department of Energy  grant
{\caps DE5--22418--281},  by the grant from the University of
Pennsylvania Research Foundation, and by the NATO research grant
 \#900--700.
I would  like to thank
the CERN Theory Division and
the Center for Theoretical Physics at Aspen, where part of
this work was done, for their hospitality.

\refout
\vfill\eject\bye